\documentclass[twocolumn,showpacs,superscriptaddress,amsmath,amssymb,prl]{revtex4}

\usepackage{graphicx}%
\usepackage{dcolumn}%

\begin{document}

\title{\textquotedblleft Waterfalls{\textquotedblright} in cuprates}

\author{A. A. Kordyuk}
\affiliation{IFW Dresden, P.O. Box 270116, D-01171 Dresden, Germany}
\affiliation{Institute of Metal Physics of National Academy of Sciences of Ukraine, 03142 Kyiv, Ukraine}

\author{S. V. Borisenko}
\author{D. S. Inosov}
\author{V. B. Zabolotnyy}
\affiliation{IFW Dresden, P.O. Box 270116, D-01171 Dresden, Germany}

\author{J.~Fink}
\affiliation{IFW Dresden, P.O. Box 270116, D-01171 Dresden, Germany}
\affiliation{BESSY GmbH, Albert-Einstein-Strasse 15, 12489 Berlin, Germany}

\author{B. B\"uchner}
\affiliation{IFW Dresden, P.O. Box 270116, D-01171 Dresden, Germany}

\author{R. Follath}
\affiliation{BESSY GmbH, Albert-Einstein-Strasse 15, 12489 Berlin, Germany}

\author{V. Hinkov}
\author{B. Keimer}
\affiliation{Max-Planck Institut f\"ur Festk\"orperforschung, 70569 Stuttgart, Germany}

\author{H. Berger}
\affiliation{Institut de Physique de la Mati\'ere Complexe, EPFL, 1015 Lausanne, Switzerland}

\date{Feb 1, 2007}%

\begin{abstract}
New hot topic in ARPES on HTSC, the observation of the so called \textquotedblleft waterfalls", is addressed. The energy scale at about 0.2-0.3 eV that can be derived from the coherent component of ARPES spectra measured along the nodal direction is not new but has been already discussed in terms of a coupling to a bosonic continuum. However, the \textquotedblleft waterfalls", namely the long vertical parts of the experimental dispersion around the center of the Brillouin zone (BZ), seem to be purely artificial. They are a consequence of simple matrix-element effect: a complete suppression of the photoemission intensity from both the coherent and \textquotedblleft incoherent" components. When the matrix-elements are taken into account, the latter reveals a grid-like structure along the bonding directions in the BZ.
\end{abstract}


\preprint{\textit{xxx}}

\maketitle

Since the mechanism of superconductivity in cuprates remains an unresolved issue, the observation of the relevant energy scale in their electronic excitation spectrum is naturally of great importance. Recently, a renewed interest to this problem has been rekindled by the finding of a \textquotedblleft high energy anomaly" in the photoemission spectra of cuprates as reported by several photoemission groups \cite{1,2,3,4,5,6}. This anomaly is explained in terms of new ideas such as a disintegration of the quasiparticles into a spinon and holon branch \cite{1}, polaronic effects induced by strong local spin correlations \cite{2}, coherence-incoherence crossover \cite{4,5}, as well as by more traditional high energy spin fluctuations \cite{3,7}. The anomaly is termed \textquotedblleft waterfalls" and is observed as extended vertical parts of the quasiparticle dispersion around the center of the Brillouin zone (BZ) (Fig.1a,b). A similar phenomenon is also found at the ($\pi$,0) point, for example (Fig.1c).

\begin{figure*}[t]
\includegraphics[width=12cm]{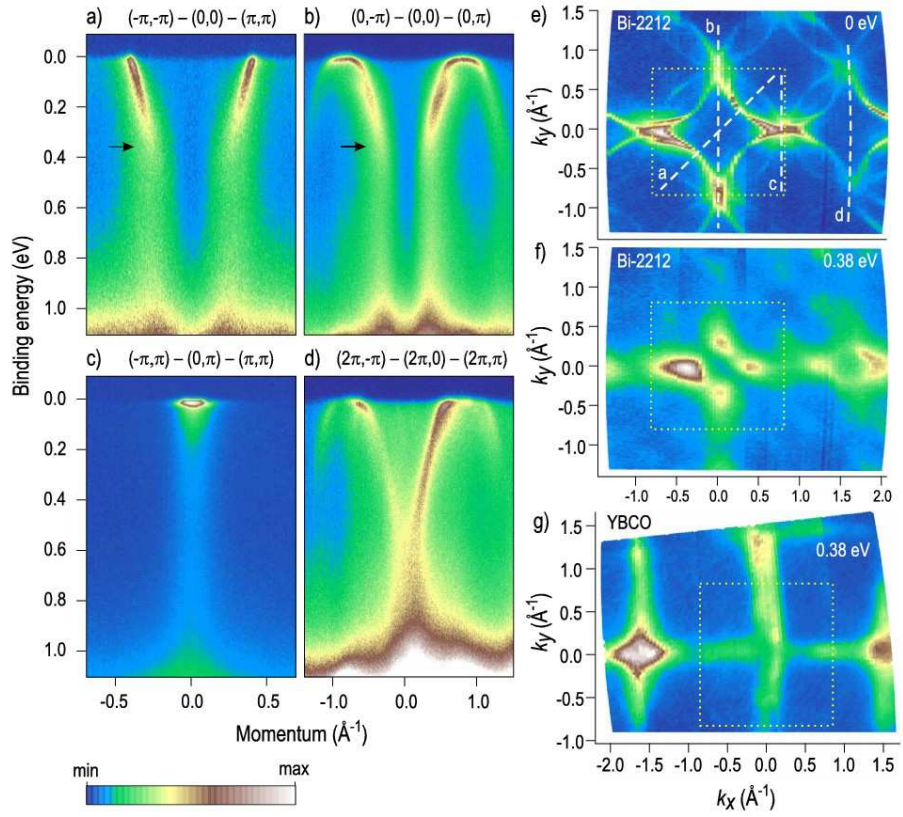}
\caption{\label{Fig1} Typical snapshots of the one-particle excitation spectrum of (Bi,Pb)2Sr2CaCu2O8 (a-f) and YBa2Cu3O6.8 (g) as seen by angle resolved photoemission. The spectra shown in panels a-d are measured along the cuts marked on the Fermi surface map (e). f, g The distribution of the spectral weight at 0.38 eV below the Fermi level for BSCCO and YBCO samples, respectively. The 1st Brillouin zone is confined by the dotted squares on the maps.}
\end{figure*}

What is anomalous here? The bend in the experimental dispersion at about 0.2-0.3 eV is a simple consequence of the renormalization maximum \cite{8} and has been already explained in terms of coupling to a continuum of bosonic excitations \cite{9}. However, the waterfalls (the long tails of the spectral weight of constant width such as seen in panels a-c between 0.4 and 0.9 eV) cannot be described within the quasiparticle approach \cite{8}, and are therefore anomalous. Here we show that the new high energy scales, obtained if one treats the waterfalls as a self-energy effect, may be wrong.

Panels b and d in Fig.1 show two photoemission spectra measured along two equivalent cuts of the reciprocal space of a single-crystalline BSCCO. Since both the electronic band structure and many body effects stay invariant to any translation by a reciprocal lattice vector, the difference between these two images comes from the photoemission matrix elements which, as a rule, strongly depend on momentum. In this case the photoemission is completely suppressed in the center of the first BZ or, more precisely, along the zone diagonals (nodal directions, see Fig.1e). One can see that both the anomalously sharp kink (black arrow) and the long vertical waterfalls in Fig.1b are caused by such a suppression. 

We stress that the waterfalls seen in panels a and b are formed because the suppression affects both the coherent part of the spectrum (the spectrum weight which can be described by the quasiparticle approach \cite{8}) and the \textquotedblleft incoherent" tails. In this respect, it is interesting that the distribution of the incoherent component in momentum is localized along the bonding directions in the reciprocal space, as one can see in Fig.1f and, more clearly, in Fig.1g for an untwinned YBCO, for which the overall picture is qualitatively the same. Evidently, this incoherent component represents a new phenomenon which should be understood. The inelastic scattering of photoelectrons can be one of its possible explanations. On the other hand, the grid-like distribution of the incoherent spectral weight in momentum may hint at the presence of a one-dimensional structure \cite{10}.

The project is part of the Forschergruppe FOR538.

\end{document}